\documentclass[aps,pre,superscriptaddress,twocolumn,balancelastpage]{revtex4-1}

\usepackage[colorlinks,bookmarks=false,citecolor=blue,linkcolor=blue,urlcolor=blue]{hyperref}
\usepackage[all]{hypcap}   

\usepackage{physics}

\usepackage{amsmath,amssymb}
\usepackage{graphicx}
\graphicspath{{figures/}{/}}

\usepackage{verbatim}
\usepackage{color}

\usepackage{placeins}  
\usepackage{flafter}     
\usepackage[T1]{fontenc}
\usepackage{lmodern}

\usepackage{color}

\newcommand{\ue}{\text{e}}
\newcommand{\ui}{\text{i}}

\newcommand{\RMTE}{random matrix model}
\newcommand{\KFIM}{KFIM}
\newcommand{\Fthree}{$C_{\text{NN}}$}
\newcommand{\Ffour}{$C_{\text{NNNN}}$ }

\begin{document}

\title{Deviations from random matrix entanglement statistics\\ for kicked quantum
chaotic spin-$1/2$ chains }

\author{Tabea Herrmann}
\affiliation{TU Dresden,
Institute of Theoretical Physics and Center for Dynamics,
 01062 Dresden, Germany}

\author{Roland Brandau}
\affiliation{TU Dresden,
Institute of Theoretical Physics and Center for Dynamics,
 01062 Dresden, Germany}

\author{Arnd B\"acker}
\affiliation{TU Dresden,
Institute of Theoretical Physics and Center for Dynamics,
 01062 Dresden, Germany}

\date{\today}

\begin{abstract}
      It is commonly expected that for quantum chaotic many body systems the statistical properties approach those of random matrices when increasing the system size. We demonstrate for various kicked spin-$1/2$ chain models that
      the average eigenstate
      entanglement indeed approaches the random matrix result.
      However, the distribution of the eigenstate entanglement differs significantly.
      While for autonomous systems such deviations are expected, they are surprising for the more scrambling kicked systems.
      Similar deviations occur in a tensor-product random matrix model with all-to-all interactions.      
      Therefore we attribute the origin of the deviations for the kicked spin-$1/2$ chain models to the tensor-product structure of the
      Hilbert spaces. As a consequence this would mean that such many body systems
      cannot be described by the standard random matrix ensembles.
\end{abstract}

\maketitle

The characterization of statistical properties of complex quantum systems
has become an important tool in many areas of physics.
Such a statistical description has started
with complex nuclei, which can be described by results from
random matrix theory (RMT) \cite{GuhMueWei1998,Meh2004}.
Even single-particle systems follow RMT if the corresponding
classical dynamics is chaotic, as stated by the famous Bohigas-Giannoni-Schmit conjecture \cite{BohGiaSch1984}.
Based on a semiclassical trace formula description \cite{Gut1990}, this
can be explained for the spectral form factor
using specific properties of periodic orbits
\cite{Ber1985,SieRic2001,Sie2002,MueHeuBraHaaAlt2004,MueHeuBraHaaAlt2005,MueHeuAltBraHaa2009}.
More recently, it has become common to call a many-body system
quantum chaotic  if its level spacing distribution
follows the RMT prediction, see e.g.\ Refs. \cite{JenSha1985,HsuAng1993,MonPoiBelSir1993,JacShe1997,AviRicBer2002,San2004,PonPapHuvAba2015,KheLazMoeSon2016,AkiWalGutGuh2016,BorLueSchKnaBlo2017,LuiBar2017b,KosLjuPro2018}.
A general expectation is that for a quantum chaotic many-body systems,
also other statistics, like eigenvector distributions
or entanglement follow the RMT results \cite{Izr1990,ZelBroFraHor1996,AleKafPolRig2016}.

For autonomous many-body systems
states at the edge of the spectrum are well known to differ from RMT \cite{EisCraPle2010}
and in particular show area law entanglement \cite{Sre1993}.
However it has recently been found, that
even the midspectrum eigenstates do not fulfil the RMT expectations.
Deviations from the RMT results are found at the level of eigenvector
distributions,
such as the components of the eigenvectors \cite{SrdProSot2021} and
the fractal dimensions \cite{BaeHaqKha2019,PauCarRodBuc2021}, but also for the
entanglement entropy \cite{VidHacBiaRig2017,VidRig2017,BeuBaeMoeHaq2018,HaqClaKha2022,KliSwiVidRig2023}.
Some of these deviations can be understood analytically \cite{Hua2019,Hua2021}. An intuitive understanding
for the deviations is based on the orthogonality of the eigenstates
which implies that the non-RMT states at the edge of the spectrum
impose for local Hamiltonians an additional
structure on the midspectrum eigenstates~\cite{HaqClaKha2022}.

The situation is much less understood for Floquet many-body systems such as kicked
systems.
These systems are more scrambling in the sense that the time evolution operator is less local
than in case of autonomous systems.
Their dynamics is described by an effective Hamiltonian
which have interactions in all
possible ranges and orders {\cite{HerKieBae2023}}.
Additionally, there is no
energy dependence of the statistical properties of eigenstates.
Therefore, the theoretical understanding of the autonomous case does not carry over to the
kicked situation.
Recently it has been shown that for kicked quantum chaotic many-body systems
spectral properties such as the spectral form factor
are excellently described by RMT at least in the thermodynamic
limit \cite{KosLjuPro2018,BerKosPro2018,ChaLucCha2018b,ChaLucCha2018a,FriChaDeCha2019,FlaBerPro2020,RoyPro2020,BerKosPro2021,GarCha2021,ChaDeCha2021,FriKie2024}.
Also for some average statistical properties of eigenstates
agreement with RMT predictions was shown \cite{AleRig2014,SieLewScaZak2023}.
In contrast, it was recently found, that for the model of random Floquet circuits, the
distribution of the entanglement entropy deviates from RMT even for large systems
\cite{RodJonKhe2024}.

However,
a detailed understanding of the entanglement behavior for
quantum chaotic kicked
systems such as
spin chains
is still lacking.
In particular, will kicked spin chains approach the
expected random matrix predictions in the thermodynamic limit
of large dimension?

In this paper we establish for the prototypical class of quantum chaotic kicked spin-chains
that the average entanglement entropy and the standard deviation
approach RMT predictions.
However, surprisingly, the distribution of the von Neumann entropy
does not approach the distribution for the corresponding RMT ensemble.
The distributions stay close but distinct, even for large system sizes,
which strongly suggests that this persists also in the thermodynamic limit.
This effect also occurs for a simple random matrix model with
all-to-all interactions.
This suggests that
the deviations are not caused by the short range
interactions in the kicked spin chains, but by the
tensor-product structure of the Hilbert spaces.

~

\textit{Quantum chaotic many body systems} --
Many body systems are commonly called quantum chaotic, if their level spacing
distribution follows the predictions from RMT \cite{JenSha1985,HsuAng1993,MonPoiBelSir1993,JacShe1997,AviRicBer2002,San2004,PonPapHuvAba2015,KheLazMoeSon2016,AkiWalGutGuh2016,BorLueSchKnaBlo2017,LuiBar2017b,KosLjuPro2018}.
In the following we restrict ourselves to predictions for the circular orthogonal ensemble (COE),
which is the appropriate
RMT ensemble for systems with time reversal invariance, such as the kicked spin
chains considered for illustrations in this paper.

For a unitary time evolution operator $U$ acting on an $N$ dimensional Hilbert space,
the eigenvalue equation reads
\begin{align}
        U|\psi_n\rangle = \ue^{\ui\varphi_n}|\psi_n\rangle\;,
        \quad \text{with}\quad
        n=1,2,\dots,N\;.
\end{align}
The eigenstates $|\psi_n\rangle$ are assumed to be normalized and the
eigenphases fulfill $\varphi_n \in [-\pi,\pi)$.
Furthermore, we choose the phases
$\{\varphi_1, \varphi_2, \dots, \varphi_N\}$ to be ordered increasingly.
The (consecutive) level spacing distribution is the
distribution of the
spacings $s_n = \frac{N}{2\pi}\left(\varphi_{n+1}-\varphi_n\right)$,
with $\varphi_{N+1} := \varphi_1 + 2\pi$.
Here the pre-factor provides the unfolding, leading to
a unit mean spacing.
In the limit $N\rightarrow \infty$
the COE result of RMT is well-described by the
Wigner distribution \cite{Wig1967}
\begin{align}
        P_{\text{COE}} (s)
                \approx \frac{\pi}{2} s
                \exp\left(-\frac{\pi}{4}s^2\right)\;.
        \label{eq:spacing_COE}
\end{align}
It is expected that for quantum chaotic many-body systems
the eigenstate entanglement follows the results predicted from
RMT \cite{Lub1978,Pag1993,Sen1996,KumPan2011}.
For a bipartite system the eigenstate entanglement can be quantified
using the von Neumann entropy. To obtain this bipartite structure
one can split the $N$--dimensional system into two subsystems
of dimension $N_1$ and $N_2$ with $N = N_1 N_2$. In the following, only
the case of $N_1 = N_2 = \sqrt{N}$ will be discussed for simplicity.
The von Neumann entropy for a state $|\psi\rangle$ is defined by
\begin{align}
        S_1 = - \text{tr}(\rho_1 \ln(\rho_1))\;,
        \label{eq:von_Neumann_entropy}
\end{align}
with $\rho_1 = \text{tr}_2(\rho)$ being the reduced density matrix of
subsystem 1, resulting from tracing out the second
subsystem from the density matrix
$\rho = |\psi\rangle\hspace*{-0.1cm}\langle\psi|$.
Unentangled states can be written as product states
$|\psi\rangle = |\psi_1\rangle\otimes|\psi_2\rangle$ and
have zero entropy, while
maximally entangled states have
$S_1^{\text{max}} = \ln\left(N_1\right)$.
The average entropy of random states from the COE
is slightly reduced. For $1 \ll N_1 \le N_2$ this entropy
is given by \cite{Pag1993},
\begin{align}
        S_1^{\text{COE,}N_1\gg1} = \ln\left(N_1\right)
                - \frac{N_1}{2 N_2}\;.
        \label{eq:entropy_COE}
\end{align}
\newcommand{\coeEntropyRef}{\cite[Eq.~(48)]{KumPan2011}}
Note that in the following we have to use the exact formula
given in Ref.~{\coeEntropyRef},
which is valid without any restrictions for $N_1$ and $N_2$.

\textit{Kicked field Ising model} --
As prototypical system to study entanglement in
many body systems, we consider
the kicked field Ising model (KFIM) which is given by
the time evolution operator \cite{Pro2002,Pro2007,BerKosPro2018,BerKosPro2019}
\begin{align}
      U_{\text{\KFIM}} =& \ue^{-\ui H^z} \ue^{-\ui H^x}\;,\quad\text{with}
      \label{eq:U_KFIM}\\
      H^z =& J \sum_{j=1}^L \sigma_j^z\sigma_{j+1}^z + \sum_{j=1}^L h_j \sigma_j^z
      \label{eq:H_z}\\
      H^x =& b \sum_{j=1}^L \sigma^x_j\;.
      \label{eq:H_KFIM_x}
\end{align}
Here $\sigma^x_j$ and $\sigma^z_j$ are the standard Pauli
spin-matrices acting on the $j$th spin
and  $L$ is the number of spins in the chain so that the dimension of the
Hilbert space is $N=2^L$.
The first part $H^z$, Eq.~\eqref{eq:H_z}, contains the nearest neighbor coupling in the $z$ component
with strength $J$ and a position dependent longitudinal field $\{h_j\}$.
The second part, Eq.~\eqref{eq:H_KFIM_x}, represents a transverse magnetic field with kicking strength
$b$ which is periodically turned on and off.
This form is equivalent to a free evolution with $H^z$~\eqref{eq:H_z}
and periodic kick described by $H^x $~\eqref{eq:H_KFIM_x}.
The longitudinal field components $\{h_j\}$ are chosen randomly from a normal
distribution with mean $\overline{h}$ and standard deviation $\sigma_h$.
We consider the self dual {\KFIM} ($J=b=\pi/4$), which is time reversal invariant and obeys another antiunitary
symmetry \cite{BraWalAkiGutGuh2020,FlaBerPro2020}. Therefore, the appropriate
RMT ensemble is the T$_+$ CRE (BD I in Cartan's classification \cite{Due2004}).
However, for the level spacing distribution and the distribution of the von Neumann entropy,
we see no numerical differences between COE and T$_+$ CRE. Thus, for simplicity,
we use COE as reference.
For numerical computations
\footnote{For the three chain models we used $J=\pi/4$, $\overline{h} = 0.6$,
$\sigma_h = \pi/4$ and $b = \pi/4$ respectively $g = \pi/4$.
For large dimensions we use the POLFED algorithm \cite{SieLewZak2020,Lui2021,SieLewScaZak2023}
with number of eigenvectors calculated $n_{\text{ev}} = \min(2^L/10, 1000)$,
dimension of Krylov space $n_{\text{cv}} = 2 n_{\text{ev}}$, a pylynomial degree of
$k = 0.95 \cdot 2^{L+1} / n_{\text{cv}}$ and a target phase $\phi_{\text{tgt}} = \pi/2$.
POLFED is used for the {\KFIM} and the {\RMTE} if $L>14$, for {\Fthree} chain and {\Ffour} chain
if $L>8$. As number of realizations we use for $L=6,8,\dots, 20$ for the {\KFIM}
$300,250,100,30,20,25,15,4$, for the {\Fthree} chain and the {\Ffour} chain
$300,250,500,100,30,25,15,4$ and for
the {\RMTE} $500,300,100,50,30,25,15,4$.
For the COE we use
$25600, 76800, 204800, 204800, 491520, 327680, 20000, 10000$ states.}
we use the polynomially filtered exact diagonalization (POLFED)
algorithm \cite{SieLewZak2020,Lui2021,SieLewScaZak2023} for long chains.

\begin{figure}
      \includegraphics{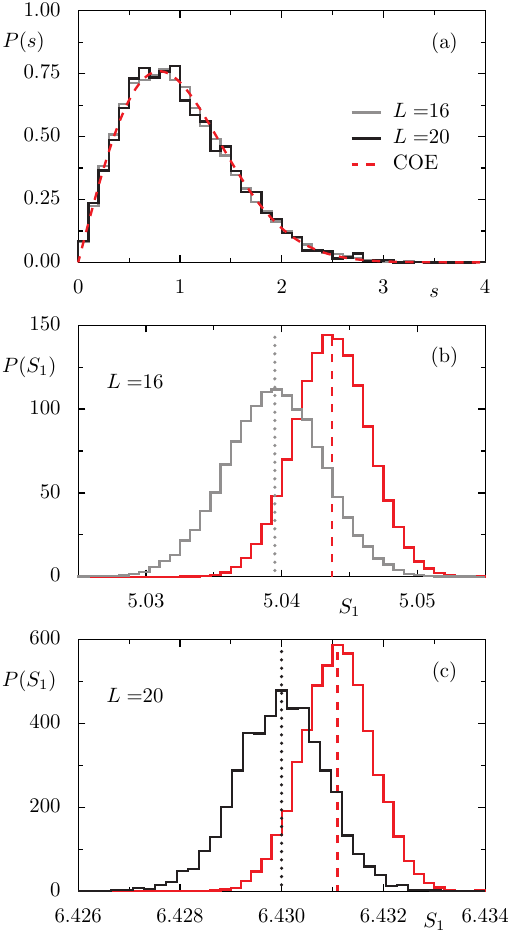}

      \caption{
      (a) Level spacing distribution for the {\KFIM} with $L=16$ (gray) and
      $L=20$ (black) and analytical predictions~\eqref{eq:spacing_COE} (red dashed line).
      (b) Distribution of von Neumann entropy for the {\KFIM} with $L=16$ (gray histogram)
      with average value shown as dotted line.
      Numerical results for random states (red histogram) and analytical prediction
      for COE (red dashed line).
      (c) similar to (b) for $L=20$, where data of {\KFIM} is shown in black.
      }
      \label{fig:kfim}
\end{figure}

Figure~\ref{fig:kfim}(a) shows the level spacing distribution for the {\KFIM} with
$L=16$ (gray) compared to the COE prediction~\eqref{eq:spacing_COE}
(red).
We see that the distribution for the {\KFIM} follows the prediction for
COE  very well. Thus, the {\KFIM} for $L=16$ qualifies as quantum chaotic for the chosen
parameters. Consequently, we would expect the von Neumann entropy of
the eigenstates to show COE behavior.

Figure~\ref{fig:kfim}(b) shows the distribution of the eigenstate entanglement
for the {\KFIM} with $L=16$ (gray) as well as the averaged von Neumann
entropy as a dotted line. The predicted average value for COE states  of
corresponding dimension {\coeEntropyRef}
is shown as a red dashed line. In lack of an analytical prediction for the distribution
of the von Neumann entropy of random states, we compare the distribution for the
{\KFIM} with numerical results for COE states.
The average von Neumann entropy for the {\KFIM} is close to the COE prediction
and also the distributions are close.
However, there are clear differences between the results for the {\KFIM} and the COE.
A possible explanation for these deviations could be that the system
size is not large enough, since we would expect the COE results to hold in the
limit of large systems sizes. Thus, we would expect that with increasing $L$
the deviations decreases and the {\KFIM} entropy approaches the COE result.

Therefore, we consider $L=20$, which also qualifies as quantum chaotic
according to the level spacing distribution,
see Fig.~\ref{fig:kfim}(a) black histogram.
Figure~\ref{fig:kfim}(c)
shows the average value and the distribution of the entropy
for the {\KFIM} with $L=20$
compared to COE results of appropriate size.
One sees, that the average values
are close to each other and that the distributions are also close.
However, compared to the results for $L=16$ in Fig.~\ref{fig:kfim}(b)
it is not clear whether the agreement between
the {\KFIM} and the COE distribution has improved,
i.e.\ whether the distributions approach each other.
Note that we have adjusted the $S_1$-axis of the plots (b) and (c)
in Fig.~\ref{fig:kfim} such that the distributions
of the different lengths can be compared visually. Thus the axis in (c) has
a much smaller $S_1$-range than in (b). This means that the absolute
distance between the average von Neumann entropy of
the {\KFIM} and the predicted
result for COE {\coeEntropyRef}, actually decreases as $L$ increases.
However, it is not possible to tell from these plots
whether the distributions approach each other or not.

\begin{figure}
      \includegraphics{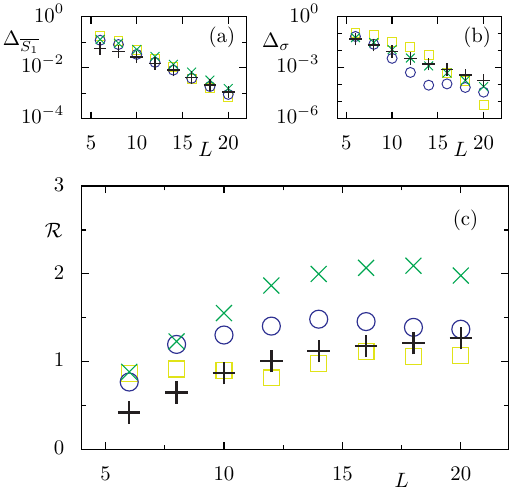}

      \caption{
      Characterization of the
      distribution of the von Neumann entropy in dependence
      on the chain length $L$ for different models.
      (a) Absolute difference between average values
          and the COE prediction, Eq.~\eqref{eq:abs_avg}.
      (b) Absolute difference between standard deviations and
          the numerically computed COE value, Eq.~\eqref{eq:abs_std}.
      (c) Ratio of the absolute difference between the average values and
          the standard deviation, Eq.~\eqref{eq:abs_avg_sigma}.
      Models: {\KFIM} (black $+$), {\Fthree} chain (green crosses), {\Ffour} chain (blue circles), and
      {\RMTE} (yellow squares).
      }
      \label{fig:entropy_over_L}
\end{figure}
\begin{figure}[b!]
      \includegraphics{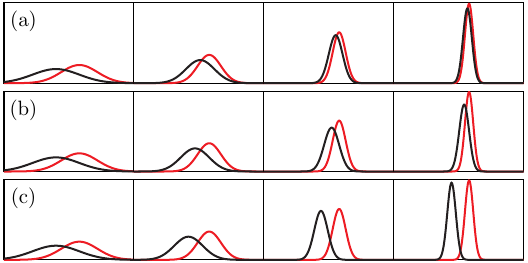}

      \caption{Visualization of different values of ratio
      $\mathcal{R}$ \eqref{eq:abs_avg_sigma}.
      As an example, we use two normal distributions for $S_1$ (black)
      and $S_1^{\text{COE}}$ (red).
      (a) Ratio continuously decreases.
      (b) Ratio stays one.
      (c) Ratio continuously increases.
      }
      \label{fig:sketch}
\end{figure}

\begin{figure}[b!]
      \includegraphics{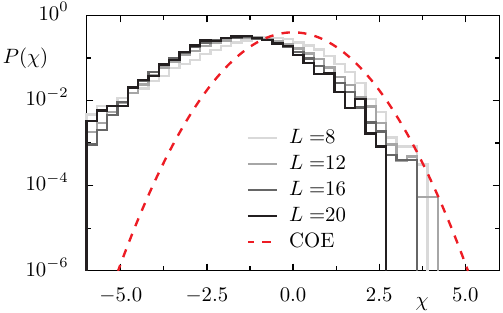}

      \caption{
      Distribution of rescaled von Neumann entropy for the {\KFIM}
      with different $L$ and
      a standard normal distribution (red dashed line)
      which provides a good approximation
      of the distribution for the COE.
      }
      \label{fig:entropy_rescaled}
\end{figure}

\textit{Approach of the distributions?} --
When two distributions approach each other, this implies
also an approach of their moments.
Figure~\ref{fig:entropy_over_L}(a) shows the absolute distance
for the first moment, i.e.\ the average of the von Neumann entropy
for the {\KFIM} and the COE {\coeEntropyRef} (black crosses), i.e.\
\begin{align}
      \Delta_{\overline{S_1}} = \left| \overline{S_1} - \overline{S_1^{\text{COE}}}\right| .
      \label{eq:abs_avg}
\end{align}
In dependence on the chain length $L$
a decrease of $\Delta_{\overline{S_1}}$ is found.
For $L>10$ it is well described by an exponential decay, as seen
by the straight line behavior in the displayed semi-logarithmic representation.
This clearly demonstrates, that the average eigenstate entanglement for the {\KFIM} indeed approaches
the COE prediction {\coeEntropyRef} in the limit of large dimensions.

In Figure~\ref{fig:entropy_over_L}(b) we see a similar behavior for the absolute
distance of the second moment, i.e.\ the standard deviation of the
distributions for {\KFIM} and COE (black crosses), i.e.\
\begin{align}
      \Delta_{{\sigma}} = \left| \sigma(S_1) - \sigma(S_1^{\text{COE}})\right|\;.
      \label{eq:abs_std}
\end{align}
Thus, also the standard deviation of the
eigenstate entanglement of the {\KFIM} approaches the COE result.

However, even if both average and standard deviation of two distributions
approach each other, it is still possible, that the distribution
remain distinct and do not approach each other:

Assuming two distributions with approaching average values and standard
deviations three different situations can arise, as illustrated in
Fig.~\ref{fig:sketch}.
Figure~\ref{fig:sketch}(a) shows the case, for which
the distributions actually approach each other.
In (b) the distributions remain close but
do not approach each other and in (c) the two
distributions form distinct peaks.
To distinguish between these three cases we
use the ratio of the absolute distance between the average
values to the standard deviation
\cite{PauCarRodBuc2021,RodJonKhe2024}
\begin{align}
      \mathcal{R}=
      \Delta_{\overline{S_1}} / \sigma(S_1)\;.
      \label{eq:abs_avg_sigma}
\end{align}
In case (a) the two distributions approach each other.
Thus this ratio is decreasing, i.e.\
the average values approach each other faster, than the decrease
of the standard deviation.
In case (b) the ratio~\eqref{eq:abs_avg_sigma} is always one,
so the distributions are always close but there is no approach.
In case (c) the ratio~\eqref{eq:abs_avg_sigma} is increasing, i.e.\
the widths of the distributions become smaller more quickly
than the averages approach each other
so that the distributions show distinct peaks.

We now apply this criterion to the {\KFIM}.
Figure.~\ref{fig:entropy_over_L}(c) shows the
ratio~\eqref{eq:abs_avg_sigma} in dependence of the chain length $L$.
Initially, the ratio~\eqref{eq:abs_avg_sigma}
increases, while for large $L$ it appears to saturate
with a value $\mathcal{R} \approx 1.25$ for $L=20$.
This leads to the remarkable conclusion, that the distribution
of the von Neumann entropy  for the {\KFIM} and the COE remain close
but do not approach each other with increasing system size.
This result is also supported by a similar behavior of the
Kullback-Leibler-divergence \cite{KulLei1951,CovTho2006} and the
Jensen-Shannon-divergence \cite{Lin1991} (not shown).
Another way to visualize the persistence of the deviations
is to consider the rescaling
$\chi = (S_1 - \overline{S_1^{\text{COE}}}) / \sigma(S_1^{\text{COE}})$.
This is shown for the {\KFIM} in comparison to the COE result
in Fig.~\ref{fig:entropy_rescaled}. One clearly sees, that 
there is no convergence towards COE.

Therefore, the question for the reason of these deviations
from the COE behavior arises.
A first guess is that these are  due to the local coupling applied
in the {\KFIM}. To test this, we discuss in the following two other kicked
spin-$1/2$ chain models and a random matrix model for kicked many-body
systems.
One could expect, that with more complex
coupling the deviations of the entropy from the COE result will decrease.

\textit{Further models --}
As second model we consider a slightly more complex coupling than the {\KFIM}.
The {\Fthree} chain is given by
the time evolution operator \cite{SieLewScaZak2023}
\begin{align}
      U_{\text{\Fthree}} =& \ue^{-\ui H^x_{\text{\Fthree}}} \ue^{-\ui H^z}\quad \text{with}
      \label{eq:U_Fthree}\\
      H^x_{\text{\Fthree}} =& \frac{g}{2} \sum_{j=1}^L \left(\sigma^x_j + \sigma^x_j \sigma^x_{j+1}\right)
      \label{eq:H_Fthree_x}
\end{align}
and $H^z$ as for the {\KFIM},
see Eq.~\eqref{eq:H_z}.
Compared to the {\KFIM}, the {\Fthree} chain has an additional nearest neighbor coupling in
the $x$ component.
The third
model is the {\Ffour} chain \cite{SieLewScaZak2023}
\begin{align}
      U_{\text{\Ffour}} =& \ue^{-\ui H^x_{\text{\Ffour}}} \ue^{-\ui H^z}\quad \text{with}
      \label{eq:U_Ffour}\\
      H^x_{\text{\Ffour}} =& \frac{g}{2} \sum_{j=1}^L \left(\sigma^x_j + \sigma^x_j \sigma^x_{j+1}
            + \frac{2}{3} \sigma^x_j \sigma^x_{j+3}\right)
      \label{eq:H_Ffour_x}
\end{align}
and $H^z$ as for the {\KFIM},
see Eq.~\eqref{eq:H_z}.
This model has an additional next--next--nearest neighbor coupling
in the $x$ component.
Note that neither the {\Fthree} chain nor the {\Ffour} chain
has a self-dual point.
Additionally, we study a {\RMTE} for quantum chaotic Floquet many
body systems given by \cite{FriKie2024},
\begin{align}
      U_{\text{\RMTE}} =&  U_c \left({U_1 \otimes U_2 \otimes \dots \otimes U_L}\right)\\
      \left[U_c\right]_{kl}
        =& \exp\left(\ui \xi_k\right)\delta_{kl}\;,
\end{align}
with $U_j$ for $j\in\{1,2,\dots,L\}$ being independent $2\times2$ matrices chosen
randomly from the CUE and
$\xi_k$ for $k \in \{1,\dots, 2^L\}$ being random variables uniformly distributed
in $\left[-\pi,\pi\right[$. Similar to the spin-$1/2$ chain models,
this model has
a Hilbert space, which is the tensor-product of $L$ two-dimensional Hilbert
spaces, but due to the all-to-all interaction lacks a spatial locality structure.
Note that due to the coupling there is also no block diagonal structure in
the resulting time evolution operator.
Numerical parameters are given in \cite{Note1}.

Figure~\ref{fig:entropy_over_L}(a) shows for each model
the absolute distance between the average von Neumann entropy
and the COE prediction~\eqref{eq:abs_avg}. All models show the same behavior,
as already found for the {\KFIM}:
Namely, for all quantum chaotic spin-$1/2$ chains and the {\RMTE} the average eigenstate entanglement approaches the COE prediction {\coeEntropyRef}.
Figure~\ref{fig:entropy_over_L}(b) demonstrates
that also the standard deviation of the von Neumann entropy approaches the COE
result for all discussed models. Figure~\ref{fig:entropy_over_L}(c) shows the
ratio~\eqref{eq:abs_avg_sigma} for all models.
In neither case the ratio becomes small when increasing $L$, but instead
approaches some constant between one and two.
Thus for all considered quantum chaotic kicked spin-$1/2$
chains as well as for the {\RMTE}, the distribution of the eigenstate
entanglement remains close to the distribution for the COE,
but the two distributions do not approach each other.

To confirm the generality of these results, we have also investigated
spin-$1/2$ chains with next-nearest-neighbor coupling and
long-range power-law coupling,
as well as the case of unequal dimensions of the subsystems and
chains with an odd number of spins (not shown). All these
examples show the same effect.
As a consequence we attribute the deviations of the entanglement
statistics  to the common feature of all these systems, which is the
tensor-product structure of the Hilbert spaces.
Note that the details of the behavior of
ratio~\eqref{eq:abs_avg_sigma} and its approach
towards saturation are system dependent.

\textit{Discussion and outlook --}
We demonstrate that for prototypical examples of quantum chaotic kicked spin-$1/2$ chains the average
eigenstate entanglement approaches the COE result when increasing the system size.
However, we find that the distribution of the entanglement entropy remains significantly
different from the distribution for the COE.
These deviations for the entropy are demonstrated for
different kicked spin-$1/2$ chain models and a random matrix model, which
takes into account the
product structure of the Hilbert spaces
of the chains but contains
all-to-all interactions.
This is in agreement with the results of Ref.~\cite{RodJonKhe2024} for the model of random Floquet circuits which also show deviations from the
full random matrix results and their origin is attributed to the locality of the model.
We have also performed numerical computations of the linear entropy (closely related to the purity),
for which similar deviations of the statistics from the results for the COE are found.

Combining these results we attribute
the deviations to the underlying product structure
of the Hilbert spaces.
As a consequence this would mean that such many body systems
cannot be described by the standard random matrix ensembles.

One could think that also other eigenvector statistics might be affected.
It turns out that the multifractal dimensions of eigenstates, see e.g.~\cite{BaeHaqKha2019,PauCarBucRod2021} and references therein,
for all considered models follow the COE results.
This means, that entanglement statistics is more sensitive to deviations from RMT than the multifractal dimensions.

In the future it would be interesting to understand the influence of the dimensions
of the subsystems.
Since we attribute the origin of the deviations to the
product structure of the Hilbert spaces, we expect to
find a similar effect also in chains with a larger spin
of the subsystems, e.g.\ spin-1 or spin-$3/2$ chains.
Another important question is whether and if so to what extent the deviations
of the eigenstate entanglement statistics from the COE
influence the dynamical behavior of the system.
Similar deviations of the entanglement of time-evolved wave-packets could mean that thermalization properties
will be different from the RMT expectations.

We thank {Felix Fritzsch, Masud Haque, Roland Ketzmerick,
Maximilian Kieler, and Ivan Khaymovich} for useful discussions.
This work was funded by the Deutsche Forschungsgemeinschaft
(DFG, German Research Foundation) --  497038782.


\end{document}